\documentclass[twocolumn]{revtex4}
\usepackage{graphicx}
\usepackage{amsmath}

\def\eqa{\begin{eqnarray}}
\def\eea{\end{eqnarray}}
\newcommand{\eq}{\begin{equation}}
\newcommand{\ee}{\end{equation}}

\newcommand{\beginsupplement}{%
        \setcounter{table}{0}
        \renewcommand{\thetable}{S\arabic{table}}%
        \setcounter{figure}{0}
        \renewcommand{\thefigure}{S\arabic{figure}}%
     }

\begin{document}

\title{Ferromagnetism and superconductivity with possible $p+ip$ pairing symmetry in partially hydrogenated graphene}

\author{Hong-Yan Lu$^{1,2*}$, Lei Hao$^{1,3}$, Rui Wang$^{1,4}$, and C. S. Ting$^{1}$}
\affiliation{$^1$ Texas Center for Superconductivity and Department of Physics,
University of Houston, Houston, Texas 77204, USA  \\$^2$ School of Physics and Electronic Information,
Huaibei Normal University, Huaibei 235000, China \\$^3$ Department of Physics, Southeast University, Nanjing 210096, China  \\$^4$ National Laboratory of Solid State Microstructures and Department of Physics, Nanjing University, Nanjing 210093, China}
%\date{\today}

\begin{abstract}
 By means of first-principles calculations, we predict two new types of partially hydrogenated graphene systems: C$_{6}$H$_{1}$ and C$_{6}$H$_{5}$, which are shown to be a ferromagnetic (FM) semimetal and a FM narrow-gap semiconductor, respectively. When properly doped, the Fermi surface of the two systems consists of an electron pocket or six hole patches in the first Brillouin zone with completely spin-polarized charge carriers. If superconductivity exists in these systems, the stable pairing symmetries are shown to be $p+ip$ for both electron- and hole- doped cases. The predicted systems may provide fascinating platforms for studying the novel properties of the coexistence of ferromagnetism and triplet-pairing superconductivity.
\end{abstract}

\maketitle

\textit{Introduction.} In recent years, hydrogenation of graphene has attracted increasing interest because it can modify the electronic and magnetic properties of graphene, providing a possible way for functioning graphene to have specially designed features. For example, fully hydrogenated graphene (graphane), with hydrogen atoms bonded to carbon atoms alternatively on both sides of the carbon plane, was theoretically predicted\cite{Sofo} and experimentally synthesized by exposing graphene in hydrogen plasma environment\cite{Elias}. From graphene to graphane, the electronic state changes from a semimetal to an insulator with a direct band gap of 3.5 eV. The hydrogenation of graphene is reversible\cite{Elias}, which provides the flexibility to manipulate the coverage of hydrogen. It is known that both graphene and graphane are nonmagnetic (NM). However, semi-hydrogenated graphene (graphone), with the hydrogen atoms on one side of graphane removed, was theoretically predicted to be a ferromagnetic (FM) semiconductor with a small indirect gap of 0.46 eV\cite{ZhouJ}. However, it was later revealed that the trigonal adsorption of hydrogen atoms in graphone is not stable, it evolves into rectangular adsorption geometry and turns into an antiferromagnetic (AFM) semiconductor with an indirect band gap of
about 2.45 eV\cite{FengL}. In addition, there are several works studying the electronic properties of graphene with various hydrogen distributions and concentrations\cite{Balog,Jaiswal,BalogJACS,Haberer,Chandrachud,YangM,GaoH, Huang,Lehtinen,Casolo,Duplock,Pujari}; it has been found that the electronic properties can be altered dramatically, e.g., opening a band gap, tuning the magnitude of the band gap of hydrogenated graphene by the hydrogen coverage, etc. Experimentally, room-temperature ferromagnetism was realized in hydrogenated epitaxial graphene\cite{Xie,Giesbers}. Exploring other kinds of hydrogenated graphene with novel properties is the main purpose of the present work.

Searching for superconductivity in graphene is another long-time pursuit. Pure graphene is not a superconductor due to the vanishing density of states (DOS) at the Dirac point. Doping graphene can bring extra electrons/holes into the system, which may give rise to superconductivity. Actually, it was theoretically predicted\cite{Profeta}, and, recently, experimentally confirmed\cite{Ludbrook} that Li-decorated graphene can dope the system with more electrons, enhance the electron-phonon coupling, and thus generate  superconductivity. It was also theoretically studied that charge doping and tensile strain also induce conventional superconductivity in graphene\cite{Liu}. Although graphane is an insulator with a large band gap, hole doping may turn it into a high-temperature electron-phonon superconductor\cite{Savini}. In terms of the possible pairing symmetry of the superconducting (SC) graphene, singlet pairing, i.e., extended $s$ wave\cite{Uchoa} or chiral $d$ wave in doped graphene (for a Review, see\cite{Black-Schaffer}), was theoretically suggested. There are also theoretical works on how to distinguish these two types of pairing symmetries\cite{Lu, Jiang}. In addition, there are also theories on possible $f$-wave triplet-pairing at certain interaction strengths in the NM phase of doped graphene\cite{Honerkamp,Kiesel,Nandkishore}. The possible existence of $p+ip$-wave triplet-pairing superconductivity in FM graphene system is an emerging issue, and so far has not been addressed.

Based on the above concerns and first-principles calculation, we show that two new hydrogenated graphene systems, C$_{6}$H$_{1}$ and C$_{6}$H$_{5}$, are, respectively, a FM semimetal and a FM semiconductor with a narrow gap of 0.7 eV separating the spin-up and the spin-down bands. We suggest that ferromagnetism and triplet-pairing superconductivity may coexist in doped C$_{6}$H$_{1}$ and C$_{6}$H$_{5}$. It is known that, the coexistence of ferromagnetism and triplet-pairing superconductivity can be realized in UGe$_{2}$\cite{Saxena,Machida}, ZrZn$_{2}$\cite{Pfleiderer}, URhGe\cite{Aoki}, and UCoGe\cite{Huy}. Therefore, C$_{6}$H$_{1}$ and C$_{6}$H$_{5}$ might provide another two fascinating platforms for studying the novel properties of the coexistence of ferromagnetism and triplet-pairing superconductivity.

\begin{figure}
\centering
\includegraphics[width=8.5cm]{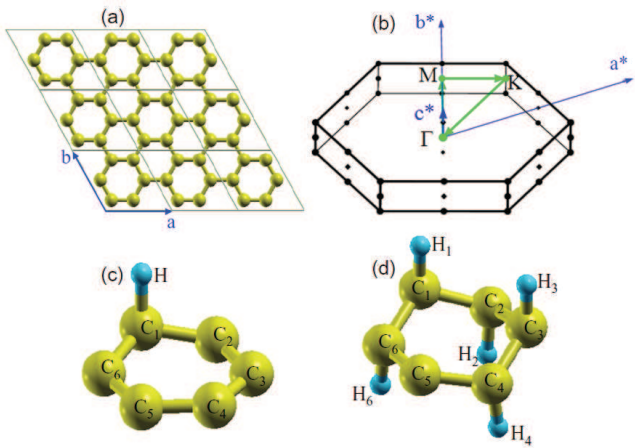}
\caption{(color online). (a) Periodic structure with six carbon atoms in a unit cell. C$_{6}$H$_{1}$ and C$_{6}$H$_{5}$ can be obtained by hydrogenating one or five carbon atoms in a unit cell. (b) The Brillouin zone and high-symmetry points for C$_{6}$H$_{1}$ and C$_{6}$H$_{5}$. (c) and (d) Optimized structure of C$_{6}$H$_{1}$ and C$_{6}$H$_{5}$ unit cells, in which the atoms are labeled to better describe their contributions.}
\end{figure}

\textit{Lattice structure and stability.} The initial lattice structures of C$_{6}$H$_{1}$ and C$_{6}$H$_{5}$ are based on the graphene lattice. We adopt a periodic structure with six carbon atoms as a unit cell, and the basis vectors are along the armchair directions of carbon atoms, as can be seen in Fig. 1(a). Graphane will be generated if all six carbon atoms in the unit cell are bonded with hydrogen atoms alternatively on both sides of the carbon plane and will further change to graphone if the hydrogen atoms on one side of the carbon plane are removed. In our case, C$_{6}$H$_{1}$ and C$_{6}$H$_{5}$ are obtained by hydrogenating one or five carbon atoms in a unit cell. The initial C-C bond length is set as 1.42 $\textmd{\AA}$ as in graphene, and the C-H bond length is set as 1.11 $\textmd{\AA}$ as in graphane\cite{Sofo}. A vacuum space of 20 $\textmd{\AA}$ normal to the graphene layer is used to avoid interactions between adjacent layers. The optimized unit-cell structure of C$_{6}$H$_{1}$ and C$_{6}$H$_{5}$ are shown in Figs. 1(c) and 1(d), respectively. The corresponding Brillouin zone (BZ) and high-symmetry points can be seen in Fig. 1(b). The calculation details and the bond lengths after relaxation are described in Supplemental Material\cite{SM}.

To prove the stability of C$_{6}$H$_{1}$ and C$_{6}$H$_{5}$, we calculate their formation energies. Using graphene and the hydrogen atom as a reference, as would be typical in the experimental setup\cite{Elias,Balog,Jaiswal,BalogJACS,Haberer}, the formation energies for C$_{6}$H$_{1}$ and C$_{6}$H$_{5}$ are -0.54 and -10.26 eV per unit cell. The negative formation energies suggest that they are thermodynamically stable. The phonon spectra for the two systems are also calculated, and there are no imaginary frequencies, indicating that they are also dynamically stable\cite{SM}.

\begin{figure}
\includegraphics[width=8.5cm]{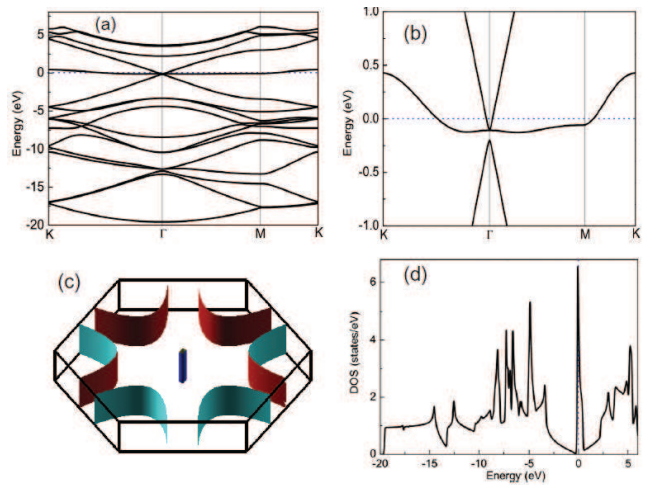}
\caption{(color online). Electronic structure of C$_{6}$H$_{1}$ in the non-spin-polarized case. (a) Electronic band structure along high-symmetry lines. The Fermi energy is set to be zero. (b) Electronic band structure around the Fermi energy. (c) FS sheets in the first BZ. (d) The total DOS.}
\end{figure}

\textit{Electronic structure of C$_{6}$H$_{1}$.} In order to check the reliability of our methods, we calculate the fully- and semi-hydrogenated cases in the 6-C unit cell. The obtained results are consistent with those reported for graphane\cite{Sofo} and graphone\cite{ZhouJ}. Then we present and discuss the numerical results for C$_{6}$H$_{1}$. We first show the results for a non-spin-polarized calculation. Fig. 2(a) shows the band structure of C$_{6}$H$_{1}$ along high-symmetry lines $K$-$\Gamma$-$M$-$K$. Compared with the band structure of graphane, which is an insulator with a large gap\cite{Sofo}, C$_{6}$H$_{1}$ is a semimetal. The most fascinating feature is that there exists a Dirac-cone-like structure with a gap of 0.23 eV at the $\Gamma$ point, as shown in Fig. 2(b) with a small energy window near the Fermi level, and there is an almost flat-band touching the bottom of the Dirac-cone band and crossing the Fermi energy away from the $\Gamma$ point. For the corresponding Fermi surface (FS), Fig. 2(c) shows that there is no dispersion along the $k_{z}$ direction, confirming the two-dimensional characteristic of the electronic structure. It contains six large hole patches around the $K$ point and one small rectangular electron pocket around $\Gamma$ point. The almost flat band will lead to a large DOS around the Fermi energy, as can be seen in Fig. 2(d). The DOS at the Fermi energy is located at a sharp Van Hove singularity peak, which indicates Stoner instability and may lead to a more stable spin-polarized state. Thus, we allow spin-polarization in our calculations for C$_{6}$H$_{1}$. The calculated total and absolute magnetizations are nearly the same, with a value of 1 $\mu_{B}$/cell, suggesting that C$_{6}$H$_{1}$ is in FM state. By comparison, we find the total energy of the FM state is 0.093 eV/cell lower than that of the NM state, indicating the FM phase is the ground state of C$_{6}$H$_{1}$.

The electronic structure of C$_{6}$H$_{1}$ in the FM state is shown in Fig. 3. From Fig. 3(a), the spin polarization of the electronic states can be clearly seen; that is, the original bands in the NM state near the Fermi level now split into spin-up and spin-down channels. The spin-up channel moves downwards, whereas the spin-down channel moves upwards. To see it more clearly,  we plot the band structure near the Fermi level in Fig. 3(b), which shows only two bands with up-spin crossing the Fermi energy. This generates one electron pocket at the $\Gamma$ point and six small hole patches at the $K$ points. The corresponding FS is plotted in Fig. 3(c). The above result confirms that C$_{6}$H$_{1}$ is a compensated semimetal, namely the electron number and the hole number are identical. From the total DOS in Fig. 3(d), we can see the relevant charge carriers near the Fermi level in the spin-polarized C$_{6}$H$_{1}$ have up-spins. Away from the Fermi level with $|E|>$ 1 eV, the bands or the electron states have very little spin polarization. Also, to find which atoms contribute the most to the spin polarization, we plot the spin-polarized and orbital-projected DOSs for all the atoms, which are presented in the Supplemental Material\cite{SM}. It is found that the spin polarization comes mainly from the 2$p_{z}$ orbitals of the C$_{2,4,6}$ atoms, as seen in Fig. 1(c), i.e., the nearest neighbours of the hydrogenated carbon atoms.

\begin{figure}
\includegraphics[width=8.5cm]{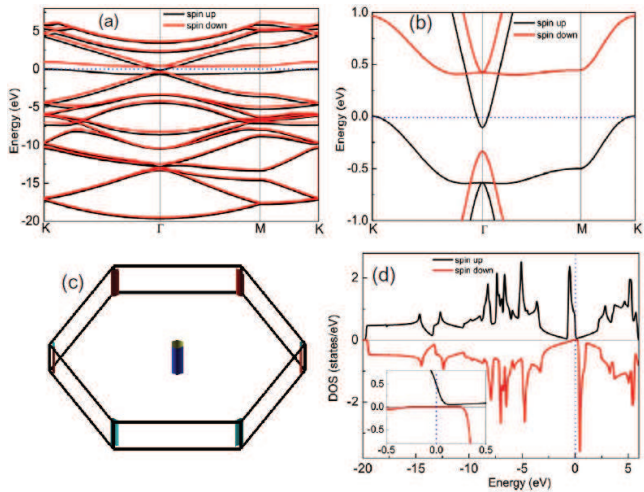}
\caption{(color online). Similar plot for C$_{6}$H$_{1}$ as in Fig. 2, but for spin-polarized calculations. Inset in (d) shows the DOS around the Fermi energy.}
\end{figure}

\textit{Electronic structure of C$_{6}$H$_{5}$.} The results are shown in Fig. 4. It is known that there is a large gap of 3.5 eV around the chemical potential for graphane. While for C$_{6}$H$_{5}$, it is equivalent to taking one hydrogen atom away from graphane (in the 6-C, 6-H unit cell), as shown in Fig. 1(d). The most obvious change in the band structure is that there is a narrow band crossing the Fermi level in the gap, as can be seen in Fig. 4(a). This almost flat band leads to a large DOS near the Fermi level [Fig. 4(b)], indicating that the NM state may be unstable against the formation of the spin-polarized state,  similar to the case for C$_{6}$H$_{1}$. Therefore, we also do a spin-polarized calculation to determine the ground state. The calculated total and absolute magnetizations are also nearly the same, with a value of 1 $\mu_{B}$/cell, which also suggests C$_{6}$H$_{5}$ is in FM state. We further check that the total energy of the FM state is 0.233 eV/cell lower than that of the NM state, indicating that the FM state is the ground state of C$_{6}$H$_{5}$.

In the FM state of C$_{6}$H$_{5}$, Fig. 4(c) shows that the largest splitting of spin-up and spin-down bands occurs near the chemical potential.  While the spin-up band moves below the chemical potential, the spin-down band increases above the chemical potential. There exists a small indirect gap of 0.7 eV separating the spin-up and the spin-down bands between the $\Gamma$ point and the $K$ points. This can also be clearly seen from the DOS in Fig. 4(d). These results suggest that C$_{6}$H$_{5}$ is a FM semiconductor with a narrow band gap. But away from the chemical potential with $|E|>$ 1.5 eV, the bands or the electron states have almost no spin polarization. Furthermore, to understand the origin of magnetism in C$_{6}$H$_{5}$, we plot the spin-polarized and orbital-projected electronic DOSs for each atom, which are also shown in the Supplemental Material\cite{SM}. The results show that the 2$p_{z}$ orbital of the C$_{5}$ atom, i.e., the unhydrogenated carbon atom, contributes the most to the DOS around the chemical potential and the spin polarization.

\begin{figure}
\includegraphics[width=8.5cm]{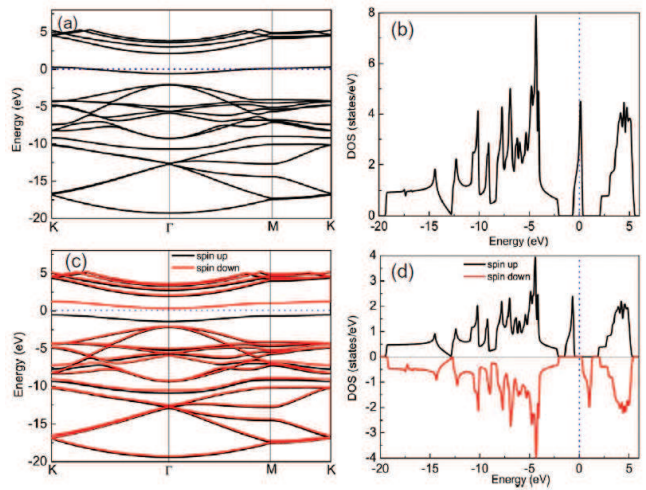}
\caption{(color online). The band structure and total DOS for C$_{6}$H$_{5}$. (a) and (b) are for non-spin-polarized calculations, while (c) and (d) are for spin-polarized calculations.}
\end{figure}

\textit{Possible $p+ip$ superconductivity in C$_{6}$H$_{1}$ and C$_{6}$H$_{5}$.} For C$_{6}$H$_{1}$, the electronic structure reveals that it is a FM semimetal, as shown in Figs. 3(a) and 3(b), with the chemical potential being set to zero. If we dope the sample with more electrons, for instance, by moving the chemical potential up to 0.25 eV, the hole pockets at the $K$ points disappear and the electron pocket at the $\Gamma$ point becomes enlarged. The corresponding FS is a cylinder around $\Gamma$, which is shown in Fig. 5(a). Similarly, if the system is doped with more holes by moving the chemical potential down to -0.25 eV, the electron pocket at the $\Gamma$ point disappears, and the hole pockets at the $K$ points get enlarged. The FS for this case is presented in Fig. 5(b). The six patches at the $K$ points are equivalent to two quasicylindrical pockets around $K_{1}$ and $K_{2}$. In both the electron- and hole-doped cases, we have only spin-up charge carriers. If there exists any superconductivity in these systems, the pairing must be triplet. The pairing mechanism is usually dominated by the $p$-wave component of the pairing interactions, which may originate from the electron-phonon and the electron-magnon couplings. We start from a most natural $p$-wave pairing interaction,
\begin{equation}
H_{p}=\sum\limits_{\mathbf{k},\mathbf{k}'}V_{p}(\mathbf{k},\mathbf{k}')
c^{\dagger}_{\mathbf{k}\uparrow}c^{\dagger}_{-\mathbf{k}\uparrow}c_{-\mathbf{k}'\uparrow}c_{\mathbf{k}'\uparrow},
\end{equation}
in which the pairing potential $V_{p}(\mathbf{k},\mathbf{k}')$ respects the full $C_{3v}$ symmetry of this material. Explicitly,
\begin{eqnarray}
V_{p}(\mathbf{k},\mathbf{k}')&=&\frac{2}{9\Omega}V^{0}_{p}[\phi_{1}(\mathbf{k})\phi^{\ast}_{1}(\mathbf{k}')   \notag \\ &&+\phi_{2}(\mathbf{k})\phi^{\ast}_{2}(\mathbf{k}')+\phi_{3}(\mathbf{k})\phi^{\ast}_{3}(\mathbf{k}')],
\end{eqnarray}
where $\Omega$ is the total area of the sample and $\phi_{i}(\mathbf{k})$ ($i=1,2,3$) are linear combinations of the basis of the twofold-degenerate $E$ representation of the $C_{3v}$ point group\cite{msd}. We have studied the leading pairing instabilities for both electron-doped and hole-doped materials.

For the electron-doped case, the pairing interaction is approximately expanded into the polynomial form as $V_{p}(\mathbf{k},\mathbf{k}')\sim k_{x}k^{'}_{x}+k_{y}k^{'}_{y}$. Define the mean-field order parameter as
\begin{equation}
\Delta(\mathbf{k})=-\sum\limits_{\mathbf{k}'}V_{p}(\mathbf{k},\mathbf{k}')\frac{\Delta(\mathbf{k}')}{E_{\mathbf{k}'}}\tanh{\frac{\beta E_{\mathbf{k}'}}{2}}=\Delta_{\alpha}\eta_\alpha(\mathbf{k}),
\end{equation}
where $\alpha$ labels different pairing channels and $\Delta_{\alpha}$ is the constant pairing amplitude. Among three pairing channels characterized by symmetry factors $\eta_1(\mathbf{k})=k_{x}$, $\eta_2(\mathbf{k})=k_{x}+k_{y}$, and $\eta_3(\mathbf{k})=k_{x}+ik_{y}$, the chiral third one is found to have the lowest ground-state energy and is thus the leading pairing instability (see the Supplementary Material for details\cite{SM}).

For the hole-doped case, the Fermi surface consists of two pockets around $K_{1}$ and $K_{2}$. It is necessary to make a comparison between interpocket BCS paring and intrapocket Fulde-Ferrell-Larkin-Ovchinnikov (FFLO) pairing\cite{FF,LO}. It turns out that the leading BCS and leading FFLO pairings, which are both of the chiral $p+ip$ form, are degenerate if the two Fermi pockets are completely circular. However, in the actual system where the two Fermi pockets have only threefold symmetry, the phase space for the FFLO pairing is suppressed compared to that for the BCS pairing. Thus, the BCS $p+ip$ pairing is the leading pairing instability\cite{SM}.

\begin{figure}
\includegraphics[width=8cm]{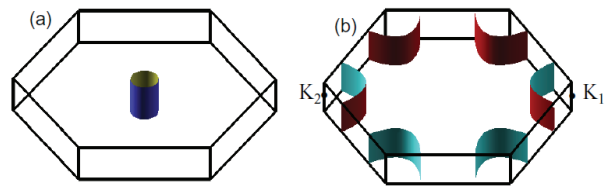}
\caption{(color online). FSs for electron- and hole-doped C$_{6}$H$_{1}$. The Fermi level moves (a) 0.25 eV upwards and (b) 0.25 eV downwards relative to the undoped case.}
\end{figure}

For C$_{6}$H$_{5}$, it is a FM semiconductor, as shown in Figs. 4(c) and 4(d). If we slightly dope it with more electrons or holes, for instance, by moving the chemical potential 0.5 eV upwards/downwards, we can obtain FSs similar to those of electron-/hole-doped C$_{6}$H$_{1}$, as shown in Figs. 5(a) and 5(b). For the electron-doped case, we have only spin-down charge carriers, and for the hole-doped case, we have only spin-up carriers. Therefore, if SC exists, it also has to show triplet-pairing symmetry. Since the C$_{6}$H$_{5}$ lattice also shows $C_{3v}$ symmetry and the FSs for doped C$_{6}$H$_{5}$ are similar to the doped C$_{6}$H$_{1}$ cases, we expect the pairing symmetries of the doped C$_{6}$H$_{5}$ are identical to those of doped C$_{6}$H$_{1}$.

\textit{Discussion and conclusion.} The above interesting theoretical result, i.e., the coexistence of ferromagnetism and superconductivity with possible $p+ip$ pairing symmetry, is expected to stimulate further experimental synthesis of these two materials. Experimentally, many kinds of hydrogenated graphene systems have been reported. For example, graphane can be fabricated by exposing graphene in hydrogen a plasma environment\cite{Elias}. More complicated systems are obtained by inducing patterned hydrogen chemisorption onto the moir\'{e} superlattice positions of graphene grown on an Ir (111) substrate\cite{Balog} or on the basal plane of graphene on a SiC substrate\cite{BalogJACS}. Moreover, Lee \textit{et al.} found that the electron beam from a scanning electron microscope can selectively remove hydrogen atoms\cite{Lee}. Most importantly, stable two-dimensional C$_{4}$H was experimentally synthesized\cite{Haberer}. There are many other investigations on various kinds of partially hydrogenated graphene systems (for review, see Ref.\cite{Review}). Therefore, based on these experimental developments, we hope the proposed C$_{6}$H$_{1}$ and C$_{6}$H$_{5}$ may also be synthesized experimentally in the future.

If the disorder of H atoms exists in the two systems, it would definitely influence their properties. However, based on the above experiments, the disorder can be maximally controlled. In view of the previous theoretical works on other similar systems, such as graphane\cite{Sofo}, graphone\cite{ZhouJ}, C$_{4}$H\cite{Haberer}, single-side-hydrogenated graphene\cite{Pujari}, doped graphane\cite{Savini}, etc., all of which were studied with ideal periodic structures, we thus consider only the periodic cases.

Doping could be achieved experimentally by gating, including using an electrolyte gate, or by charge transfer, as done in graphene\cite{Das, Gierz, Jung}. For graphane, hole doping can be obtained by partially substituting carbon atoms with boron atoms\cite{Savini}. It showed that the band structure and the DOS of graphane near the chemical potential inside and outside the band gap are practically unchanged even up to a 12.5\% boron doping. This justifies the use of a rigid-band approximation to simulate substitutional doping in graphane. The above work demonstrated that small substitutional dopings shift only the chemical potential and do not change the band structure near the chemical potential. Our current doping is less than 1.6\% hole or electron doping, which will also not change the band structure. Of course, even if the dopants are randomly distributed, they should have negligible effect on the band structure at such a small concentration.

Finally, we would like to make a comparison between the triplet-pairing superconductivity in C$_{6}$H$_{1}$/C$_{6}$H$_{5}$ systems and those in heavy-metal (UGe$_{2}$ \cite{Saxena,Machida}, URhGe \cite{Aoki}, and UCoGe \cite{Huy}) or transition-metal based metallic compounds. All of them show the coexistence of triplet-pairing superconductivity and itinerant electron ferromagnetism but originate from different electron orbitals. For uranium-based systems, ferromagnetism and the superconductivity are determined by the U 5$f$ orbitals with possible strong spin-orbital couplings. However, for transition-metal-based system ZrZn$_{2}$\cite{Pfleiderer}, Zr 4$d$ orbital electrons play the most important role. Therefore, the predicted C$_{6}$H$_{1}$/C$_{6}$H$_{5}$ may provide new platforms to study the coexistence of ferromagnetism and triplet-pairing superconductivity within $p$ electron orbitals.

In conclusion, we predicted two new types of hydrogenated graphene, C$_{6}$H$_{1}$ and C$_{6}$H$_{5}$, and found they are a FM semimetal and a FM semiconductor with a narrow gap, respectively. For doped C$_{6}$H$_{1}$ and C$_{6}$H$_{5}$, there may exist superconductivity with chiral $p+ip$ pairing symmetry, which is known to support chiral edge states and vortex zero modes, both known as Majorana fermions\cite{Read,Ivanov}. Thus, the predicted superconducting phases for C$_{6}$H$_{1}$ and C$_{6}$H$_{5}$ may provide new platforms for studying the novel physics in topological quantum computations\cite{tqc}.

We are thankful for helpful discussions with J.-X. Zhu, V. G. Hadjiev, S.-Y. Lin and R. Thomale. This work is supported by the Texas Center for Superconductivity at the University of Houston and the Robert A. Welch Foundation (Grant No. E-1146), the National Natural Science Foundation of China (Grants No. 11574108, 11104099, and 11204035), the Natural Science Foundation of Anhui Province (Grant No. 1408085QA12), and the Natural Science Research Project of Higher Education Institutions of Anhui Province (Grant No. KJ2015A120). The numerical calculations were performed at the Center of Advanced Computing and Data Systems at the University of Houston.

\vspace{0.5cm}$^*$e-mail: luhongyan2006@gmail.com

\beginsupplement
\textit{SUPPLEMENTARY MATERIAL}
\section{Computational methods}

The calculations are performed within density functional theory (DFT) as implemented in the Quantum-Espresso (QE) program\cite{QE}. We adopt Projector Augmented-Wave method\cite{PAW} to model the electron-ion interactions, and generalized gradient approximation (GGA) with Perdew-Burke-Ernzerhof parametrization\cite{PBE} for the exchange correlation potentials. The lattice dynamics is performed within the framework of the density functional perturbation theory (DFPT)\cite{Baroni} as implemented in QE\cite{QE}.

We perform a full structural optimization including both the lattice parameters and the atom positions. The precision for the convergence of total energy and force are 10$^{-7}$ Ry and 10$^{-6}$ Ry/Bohr, respectively. The cutoff for wave functions and charge density are 80 Ry and 480 Ry, respectively. The electronic integration is performed over a 12$\times$12$\times$1 k-point mesh. For Fermi surface (FS) and DOS calculations, denser 48$\times$48$\times$4 and 120$\times$120$\times$1 k-point grids are respectively adopted. For the phonon calculation, the dynamical matrices are calculated on a 6$\times$6$\times$1 q-point grid.

\section{Bond lengths for C$_{6}$H$_{1}$ and C$_{6}$H$_{5}$}

For the two systems, after structural relaxation, because of the formation of C-H bonds, the hydrogenated carbon atoms are pulled out of the graphene plane and the hybridization of them changes from $sp^{2}$ to $sp^{3}$, which is similar to graphane and graphone. For C$_{6}$H$_{1}$, the C-C (with one C bonded with H) and C-H bond lengths are 1.496 and 1.137 $\textmd{\AA}$, respectively. Whereas for the unhydrogenated carbon atoms, the C-C bond length barely changes. For C$_{6}$H$_{5}$, the C-H bond length with the H atoms above and below the graphene plane are 1.104 and 1.131 $\textmd{\AA}$, respectively. The C-C bond lengths are 1.537 $\textmd{\AA}$ (both C are bonded with H) and 1.485 $\textmd{\AA}$ (one C is bonded with H).

\section{Phonon spectra for C$_{6}$H$_{1}$ and C$_{6}$H$_{5}$}

We calculate the phonon spectra for the two systems, and the results are shown in Fig. 1. There is a wide range of frequency, extending to about 2590 and 2950 cm$^{-1}$ for C$_{6}$H$_{1}$ and C$_{6}$H$_{5}$, respectively. It is seen that there are no imaginary frequencies for them, indicating that they are dynamically stable.

\begin{figure}
\includegraphics[width=8.5cm]{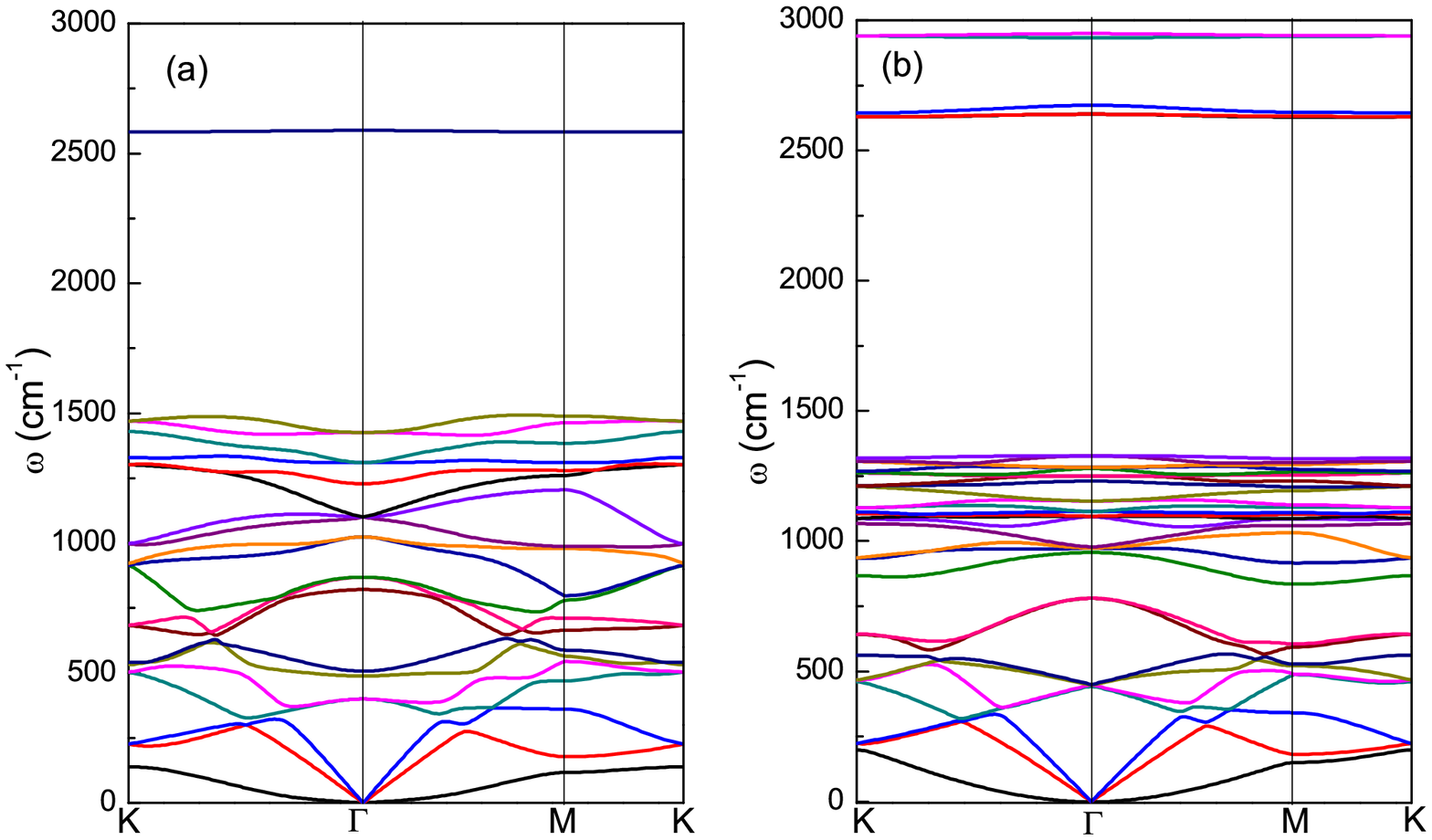}
\caption{Phonon spectra for (a) C$_{6}$H$_{1}$, and (b) C$_{6}$H$_{5}$.}
\end{figure}

\section{Spin-polarized and orbital-projected electronic DOS for C$_{6}$H$_{1}$ and C$_{6}$H$_{5}$}

For C$_{6}$H$_{1}$ and C$_{6}$H$_{5}$, to find which atoms contribute most to the spin polarization, we plot the spin-polarized and orbital-projected DOS for all the atoms. The results for the two systems are shown in Fig. 2 and Fig. 3, respectively. We first discuss the result of C$_{6}$H$_{1}$. Combined with Fig. 1(c) in the main text, it is  found that the 7 atoms in a unit cell can be divided into four groups: (1) H atom, (2) C$_1$ atom, (3) C$_{2,4,6}$ atoms, and (4) C$_{3,5}$ atoms, with each group showing the same DOS. The reason is, by building the supercell of C$_{6}$H$_{1}$, it is easy to see that C$_{2}$, C$_{4}$, and C$_{6}$ atoms are the nearest neighbours of the hydrogenated carbon atoms. Therefore, the lattice symmetry determines that the DOS of them are the same. Similarly, C$_{3}$ and C$_{5}$ atoms are the next nearest neighbours of the hydrogenated carbon atoms, and also show the same DOS. From Fig. 2, we can see that C$_{2,4,6}$ atoms, i.e., the nearest neighbours of the hydrogenated carbon atoms contribute most to the total DOS around the Fermi energy and also to the spin polarization. The next main contribution is from the H atom, while the contributions of other atoms can be ignored. To be more specific, we check and find that the spin polarization mainly comes from the 2p$_{z}$ orbital ($\pi$ electrons) of the C$_{2,4,6}$ atoms. This can be understood that the $sp^{2}$ hybridized C$_1$ atom becomes $sp^{3}$ hybridized after hydrogenation and forms strong $\sigma$ bonds with H atom, while the $\pi$ electrons of unhydrogenated C$_{2,4,6}$ atoms are more localized compared with those in pure graphene, and therefore become spin polarized. However, for the next nearest neighbours C$_{3,5}$ atoms, the $\pi$ electrons are still delocalized and therefore contribute little to spin polarization.

For C$_{6}$H$_{5}$, we plot the spin-polarized and orbital-projected electronic DOS for each atom, which are shown in Fig. 3. According to the symmetry of the lattice, the 11 atoms in the unit cell (Fig. 1(d) in the main text) can be divided into 5 groups: (1) H$_{1,3}$ atoms, (2) H$_{2,4,6}$ atoms, (3) C$_{1,3}$ atoms, (4) C$_{2,4,6}$ atoms, and (5) C$_{5}$ atom, with each group showing the same DOS. The calculated results in Fig. 3 show that the 2p$_{z}$ orbital ($\pi$ electron) of C$_{5}$ atom, i.e., the unhydrogenated carbon atom contributes the most to the DOS around the chemical potential and the spin-polarization, whereas the 1$s$ orbitals of H$_{2,4,6}$ atoms, i.e., the hydrogen atoms on the nearest neighbours of the unhydrogenated carbon atom contribute very little. For other atoms, because of the formation of $sp^{3}$ hybridization after hydrogenation, they do not contribute to the DOS and spin-polarization near the chemical potential.

\begin{figure}
\includegraphics[width=8.5 cm]{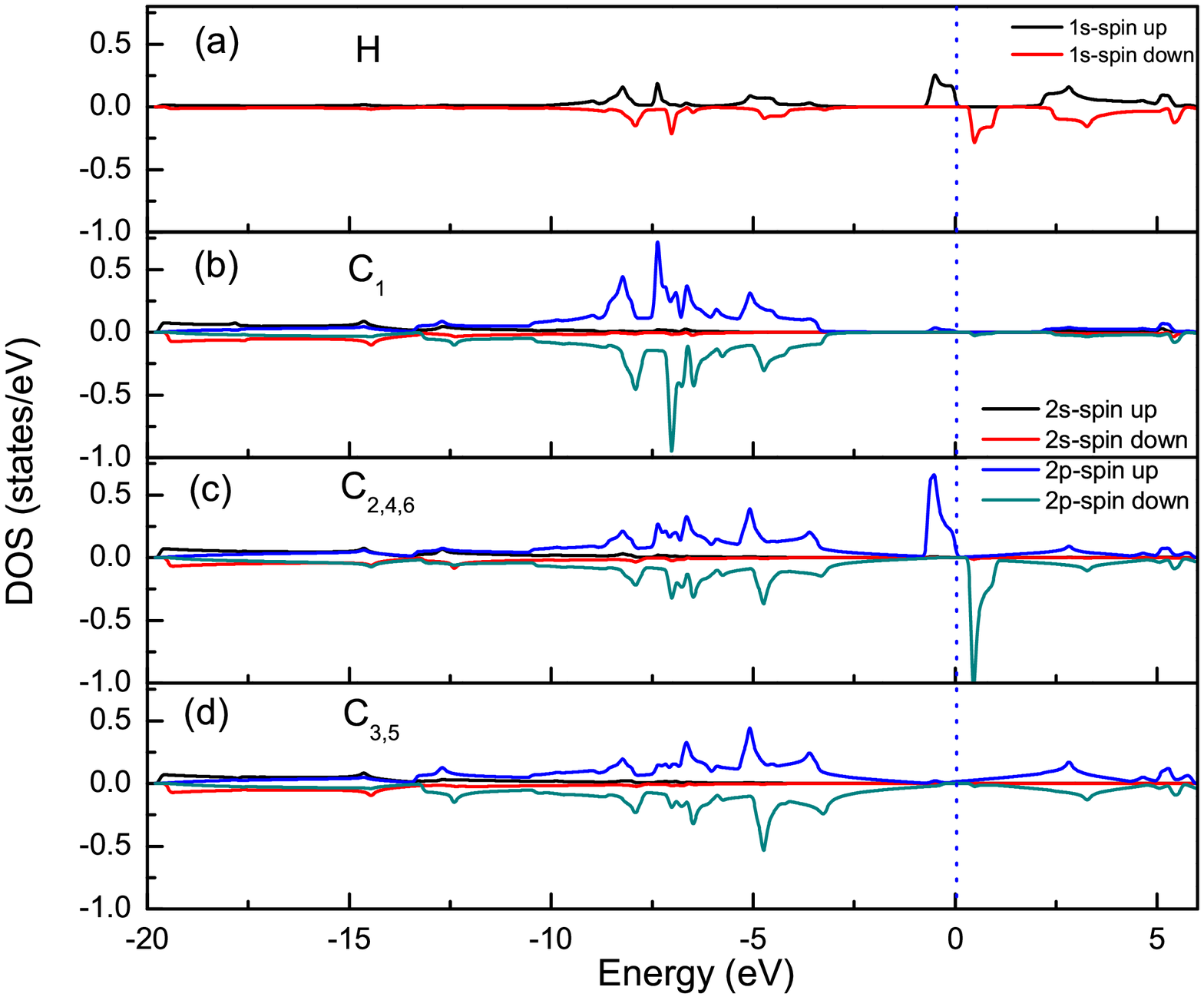}
\caption{Spin-polarized and orbital-projected electronic DOS for (a) H atom, (b) C$_1$ atom, (c) C$_{2,4,6}$ atoms, and (d) C$_{3,5}$ atoms in C$_{6}$H$_{1}$.}
\end{figure}

\begin{figure}
\includegraphics[width=8.5 cm]{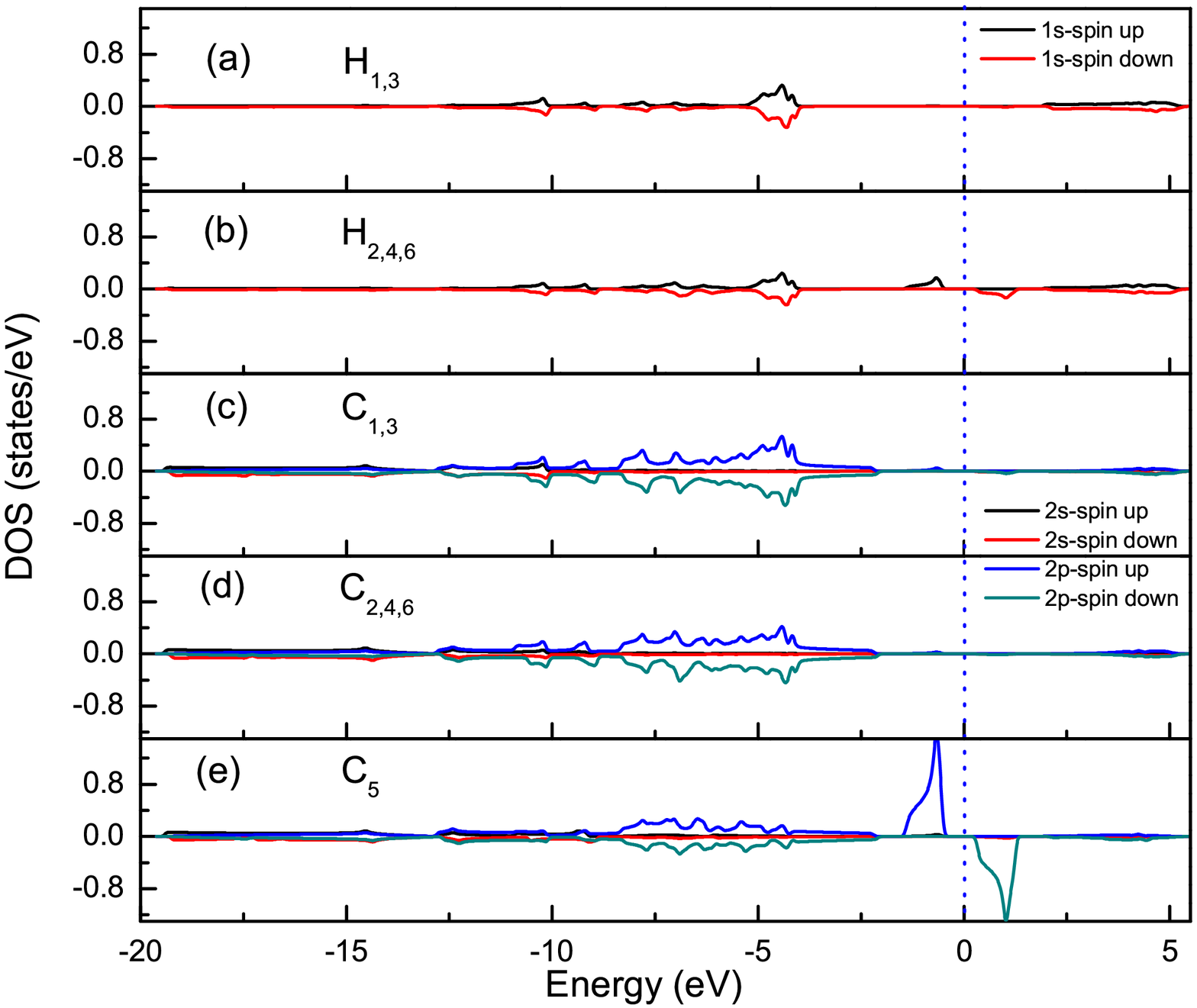}
\caption{Spin-polarized and orbital-projected electronic DOS for (a) H$_{1,3}$ atoms, (b) H$_{2,4,6}$ atoms, (c) C$_{1,3}$ atoms, (d) C$_{2,4,6}$ atoms, and (e) C$_{5}$ atom in C$_{6}$H$_{5}$.}
\end{figure}

\section{Superconductivity instabilities}

As was explained in the main text, by doping electrons (holes) to C$_{6}$H$_{1}$ and C$_{6}$H$_{5}$, we can get a fully spin-polarized Fermi surface consisting of one (two) circular electron (hole) pocket(s) at the $\Gamma$ point (two $K$ points) of the Brillouin zone. While the spin-polarization of the charge carriers for electron-doped and hole-doped C$_{6}$H$_{5}$ are opposite, they are the same for C$_{6}$H$_{1}$.

As an important low-temperature instability to all metallic systems, it is interesting to study the possible superconducting transition in these new ferromagnetic metals. This is particularly true for C$_{6}$H$_{1}$, since it is much easier to dope charge carriers to this semimetal. We thus focus in what follows on C$_{6}$H$_{1}$, the conclusions for which should also be applicable to C$_{6}$H$_{5}$.

Since we have fully spin-polarized Fermi surfaces for both electron-doped and hole-doped cases, all possible pairing channels should be equal-spin spin-triplet. To conform to the Fermi statistics, the spatial part of the superconducting order parameter should be an odd function. As a result, while the full pairing interaction might have both even-parity and odd-parity components, only the odd-parity component is active in inducing a superconducting transition.

We first consider the electron-doped case, for which the Fermi surface is a circle around the $\Gamma$ point (see Fig. 5(a) in the main text). To gain a qualitative understanding of the possible pairing channel, we consider a generic pairing interaction in the $p$-wave channel, which is the leading odd-parity pairing interaction. We assume that the pairing interaction is nonzero only within a small window of $[-\omega_{c},\omega_{c}]$ around the Fermi surface. We thus write the phenomenological pairing interaction as
\begin{equation}
H_{p}=\sum\limits_{\mathbf{k},\mathbf{k}'}V_{p}(\mathbf{k},\mathbf{k}')
c^{\dagger}_{\mathbf{k}\uparrow}c^{\dagger}_{-\mathbf{k}\uparrow}c_{-\mathbf{k}'\uparrow}c_{\mathbf{k}'\uparrow},
\end{equation}
where the summation over $\mathbf{k}$ and $\mathbf{k}'$ are restricted to be close to the chemical potential. The lattice constant $a=3a_{C-C}$ will be set as the length unit (i.e., $a=1$) in the following analysis. To conform to the $C_{3v}$ symmetry of the C$_{6}$H$_{1}$ and C$_{6}$H$_{5}$ lattices, the pairing potential is taken to be the following form
\begin{eqnarray}
V_{p}(\mathbf{k},\mathbf{k}')&=&\frac{2}{9\Omega}V^{0}_{p}[\phi_{1}(\mathbf{k})\phi^{\ast}_{1}(\mathbf{k}')   \notag \\ &&+\phi_{2}(\mathbf{k})\phi^{\ast}_{2}(\mathbf{k}')+\phi_{3}(\mathbf{k})\phi^{\ast}_{3}(\mathbf{k}')],
\end{eqnarray}
where $\Omega$ is the total area of the sample. $\phi_{i}(\mathbf{k})$ ($i=1,2,3$) are linear combinations of the basis of the twofold-degenerate $E$ representation of the $C_{3v}$ point group, and are defined as $\phi_{1}(\mathbf{k})=e^{i\mathbf{k}\cdot\mathbf{a}_{1}}-e^{i\mathbf{k}\cdot\mathbf{a}_{2}}$, $\phi_{2}(\mathbf{k})=e^{i\mathbf{k}\cdot\mathbf{a}_{2}}-e^{i\mathbf{k}\cdot\mathbf{a}_{3}}$, and $\phi_{3}(\mathbf{k})=-(\phi_{1}(\mathbf{k})+\phi_{2}(\mathbf{k}))= e^{i\mathbf{k}\cdot\mathbf{a}_{3}}-e^{i\mathbf{k}\cdot\mathbf{a}_{1}}$,\cite{msd} in which $\mathbf{a}_{1}=(\frac{1}{2},\frac{\sqrt{3}}{2})a$, $\mathbf{a}_{2}=(\frac{1}{2},-\frac{\sqrt{3}}{2})a$, and $\mathbf{a}_{3}=-(\mathbf{a}_{1}+\mathbf{a}_{2})$. For the present electron-doped system, the Fermi surface of which is close to the center of the Brillouin zone (BZ) (i.e., the $\Gamma$ point), it is reasonable to express the physical quantities in terms of polynomials of the wave vector.  Thus, we expand the pairing potential approximately as
\begin{equation}
V_{p}(\mathbf{k},\mathbf{k}')=\frac{1}{\Omega}V^{0}_{p}(k_{x}k^{'}_{x}+k_{y}k^{'}_{y})
=\frac{1}{\Omega}V^{0}_{p}kk^{'}\cos(\theta-\theta'),
\end{equation}
where $k$ ($k'$) and $\theta$ ($\theta'$) are the radial and azimuthal coordinate components of $\mathbf{k}$ ($\mathbf{k}'$) in the polar coordinate. Define $\xi_{\mathbf{k}}=k^{2}/2m-\mu$ as the energy of electrons in the spin-up band relative to the chemical potential $\mu$, $V^{0}_{p}$ is taken as a nonzero constant if both $\xi_\mathbf{k}$ and $\xi_\mathbf{k}'$ lie within the range of $[-\omega_{c},\omega_{c}]$.

Introducing a mean-field decoupling to the pairing interaction by defining the superconducting order parameter
\begin{equation}
\Delta(\mathbf{k})=\sum\limits_{\mathbf{k}'}V_{p}(\mathbf{k},\mathbf{k}')<c_{-\mathbf{k}'\uparrow}c_{\mathbf{k}'\uparrow}>,
\end{equation}
we arrive at a mean-field Hamiltonian of the form
\begin{eqnarray}
H_{MF}&=&\frac{1}{2}\sum\limits_{\mathbf{k}}\psi^{\dagger}_{\mathbf{k}}\begin{pmatrix} \xi_\mathbf{k} & 2\Delta(\mathbf{k}) \\
2\Delta^{\ast}(\mathbf{k}) & -\xi_{-\mathbf{k}}
\end{pmatrix}\psi_{\mathbf{k}}  \notag \\ &&-\sum\limits_{\mathbf{k}}\Delta(\mathbf{k})<c^{\dagger}_{\mathbf{k}\uparrow}c^{\dagger}_{-\mathbf{k}\uparrow}>,
\end{eqnarray}
where $\psi^{\dagger}_{\mathbf{k}}=[c^{\dagger}_{\mathbf{k}\uparrow},c_{-\mathbf{k}\uparrow}]$ is the Nambu basis for the spin-polarized band crossing the chemical potential. The $1/2$ factor in front of the $\mathbf{k}$ summation removes the particle-hole redundancy in the Nambu representation. The self-consistent equation for the pairing order parameter is obtained from the definition as
\begin{equation}
\Delta(\mathbf{k})=-\sum\limits_{\mathbf{k}'}V_{p}(\mathbf{k},\mathbf{k}')\frac{\Delta(\mathbf{k}')}{E_{\mathbf{k}'}}\tanh{\frac{\beta E_{\mathbf{k}'}}{2}},
\end{equation}
where $\beta=1/(k_{B}T)$ is the inverse temperature and $E_{\mathbf{k}}=\sqrt{\xi_\mathbf{k}^{2}+4|\Delta(\mathbf{k})|^2}$ is the quasiparticle energy.

Among all the possible $p$ wave equal-spin triplet pairing channels, whose symmetry factors $\eta(\mathbf{k})$ are linear combinations of $k_x$ and $k_y$, we make a comparison over three most probable channels: (1) $\eta_1(\mathbf{k})=k_{x}$, (2) $\eta_2(\mathbf{k})=k_{x}+k_{y}$, and (3) $\eta_3(\mathbf{k})=k_{x}+ik_{y}$. This choice is motivated from the three equivalent representations of the pairing potential: $k_{x}k^{'}_{x}+k_{y}k^{'}_{y}=\frac{1}{2}[(k_{x}+k_{y})(k^{'}_{x}+k^{'}_{y})+(k_{x}-k_{y})(k^{'}_{x}-k^{'}_{y})] =\frac{1}{2}[(k_{x}+ik_{y})(k^{'}_{x}-ik^{'}_{y})+(k_{x}-ik_{y})(k^{'}_{x}+ik^{'}_{y})]$. The three pairings are all nonunitary.\cite{machida01,samokhin02} We thus have for the $\alpha$-th ($\alpha=$1,2,3) pairing channel the ansatz
\begin{equation}
\Delta_{\alpha}(\mathbf{k})=\Delta_{\alpha}\eta_\alpha(\mathbf{k}),
\end{equation}
where $\Delta_{\alpha}$ is a real number measuring the pairing amplitude in the $\alpha$-th channel. Put this ansatz for $\Delta_{\alpha}(\mathbf{k})$ back to Eq.(6), we get the following equation for determining the pairing amplitude $\Delta_{\alpha}$ and the corresponding superconducting transition temperature
\begin{equation}
\sum\limits_{\mathbf{k}}|\eta_{\alpha}(\mathbf{k})|^{2}
=-\sum\limits_{\mathbf{k}\mathbf{k}'}V_{p}(\mathbf{k},\mathbf{k}')\eta^{\ast}_{\alpha}(\mathbf{k})\eta_{\alpha}(\mathbf{k}')\frac{\tanh\frac{\beta E_{\mathbf{k}'}}{2}}{E_{\mathbf{k}'}},
\end{equation}
where the summations over $\mathbf{k}$ and $\mathbf{k}'$ are restricted by $|\xi_{\mathbf{k}}|\le \omega_{c}$ and $|\xi_{\mathbf{k}'}|\le \omega_{c}$.

It is easy to see from Eq.(8) that, all the three pairing channels considered have the same superconducting transition temperature
\begin{equation}
k_{B}T_{C}\simeq 1.14\omega_{c}e^{\frac{\pi}{V^{0}_{p}m^{2}\mu}}=1.14\omega_{c}e^{\frac{1}{V^{0}_{p}N(\mu)k^{2}_{F}}},
\end{equation}
where $N(\mu)=m/(2\pi)$ is the normal state density of states for the fully spin-polarized parabolic band and $k_{F}=\sqrt{2m\mu}$. In the limit of weak coupling, for which the pairing amplitude is small compared to the energy scales ($\mu$ and $\omega_{c}$) of the problem, we can also get an analytical solution of the pairing amplitude for $\alpha=3$ at zero temperature
\begin{equation}
\Delta_{3}\simeq \frac{\omega_{c}}{k_{F}}e^{\frac{1}{V^{0}_{p}N(\mu)k^{2}_{F}}}.
\end{equation}
From Eqs.(9) and (10), we see that the relevant energy scale determining the superconducting transition is $V^{0}_{p}N(\mu)k^{2}_{F}=V^{0}_{p}m^{2}\mu/\pi$.

To identify the leading pairing instability, we have to compare their ground state energies. The ground state energy is defined as the zero temperature average of the mean-field Hamiltonian. Since pairing occurs only within a small energy window around the chemical potential, it is enough to compare the energy for this part of the spin-polarized band. Denoting the part of the ground state energy relevant to superconducting pairing as $E^{FS(\alpha)}_{GS}$ ($\alpha$=1,2,3), we have
\begin{equation}
E^{FS(\alpha)}_{GS}/\Omega=\int\frac{d^{2}\mathbf{k}}{(2\pi)^{2}}\frac{2|\Delta_{\alpha}(\mathbf{k})|^{2}-E^{2}_{\mathbf{k}}}{2E_{\mathbf{k}}} \theta(\xi_{\mathbf{k}}+\omega_{c})\theta(\omega_{c}-\xi_{\mathbf{k}}).
\end{equation}
$\theta(x)$ is the Heaviside step function, which is one for $x\ge0$ and zero otherwise.

For a typical set of parameters ($m=0.5$, $\mu=0.5$, $\omega_{c}=0.1$), the pairing amplitudes and averaged ground state energies are shown separately in Fig. 4(a) and Fig. 4(b), as a function of the effective strength of the pairing potential. As is clear from the figures, the two nodal phases with symmetry factors $\eta_{1}(\mathbf{k})$ and $\eta_{2}(\mathbf{k})$ are degenerate in energy. The ground state is the fully-gapped phase with symmetry factor $\eta_{3}(\mathbf{k})$. This stable phase, which is chiral in nature, is well-known to support chiral edge states and vortex zero modes both known as Majorana fermions.\cite{read00,ivanov01}

\begin{figure}
\centering
\includegraphics[width=7cm]{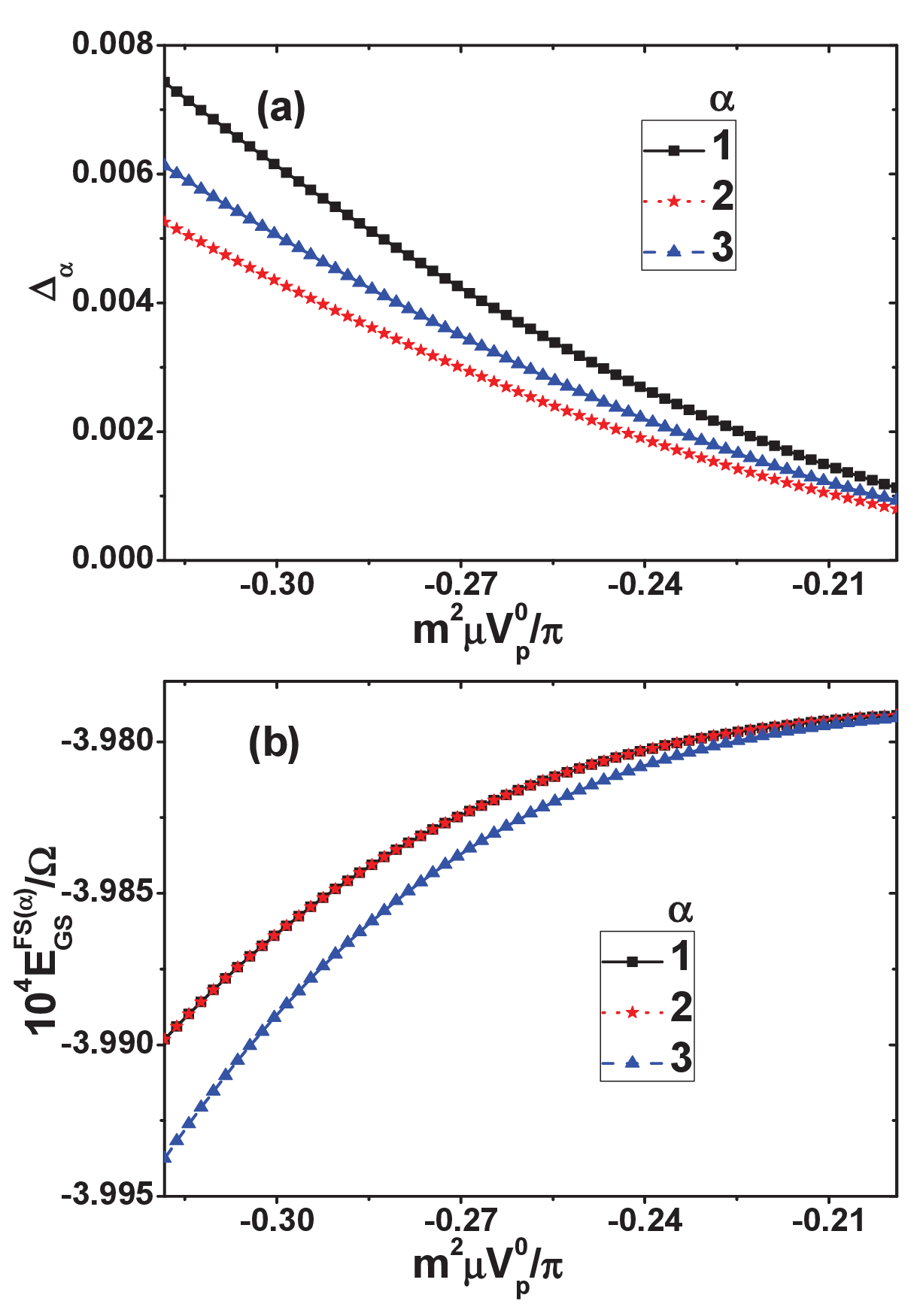}
\caption{(a) Mean-field pairing amplitudes and (b) averaged ground state energies (enlarged by a factor of $10^4$), as a function of the effective pairing strength $m^2\mu V^{0}_{p}/\pi$. The three pairings are defined in the text. For each parameter considered, $\Delta_{1}$ is exactly $\sqrt{2}$ times of $\Delta_{2}$. Only those states participating in pairing are included in calculating the ground state energy.}
\end{figure}

Now, we proceed to study the hole-doped case. As shown in Fig. 5(b)  of the main text, the Fermi surface consists of six disconnected patches at the $K$ points. From the periodicity in the momentum space, the Fermi surface can also be regarded equivalently as consisting of two closed pockets situating respectively at the two inequivalent $K$ points at $\mathbf{K_{1}}=(1,0)\frac{4\pi}{3a}$ and $\mathbf{K_{2}}=(-1,0)\frac{4\pi}{3a}=-\mathbf{K_{1}}$, where $a=3a_{C-C}$ has been set as the length unit. It should be noticed that, the effective mass of the charge carriers for the hole-doped case is much larger than that for the electron-doped case, thus the superconducting transition temperature for hole-doped case might also be significantly larger than that for the electron-doped case.

Because the Fermi surface and nearby low energy states are also fully spin-polarized, the pairing instability can only realize in the odd-parity spin-triplet channel, of which the $p$-wave component is usually dominant. Therefore, it is reasonable to start from the same phenomenological pairing potential, Eq.(2), used for the electron-doped case.

Since we now have two disconnected Fermi pockets situated symmetrically with respect to the $\Gamma$ point, the pairing instability can occur either in the inter-pocket channel or in the intra-pocket channel. While the former is the conventional Bardeen-Cooper-Schrieffer (BCS) state \cite{bcs}, the latter case is the Fulde-Ferrell-Larkin-Ovchinnikov (FFLO) state with nonzero center of mass momentum \cite{fflo}. Now, it is suitable to comment on the symmetry of the Fermi surface. While for samples with extremely low doping concentrations the Fermi pockets surrounding the $K$ points are almost circular for C$_{6}$H$_{1}$, they show increasingly larger anisotropy as the doping concentration increases. The deviation from circular Fermi pocket is also more obvious in C$_{6}$H$_{5}$ than for C$_{6}$H$_{1}$. However, to make possible an analytical analysis, we will focus in what follows on Fermi pockets having perfect circular symmetry centering at each $K$ point. The implications of the anisotropy in the Fermi pockets are discussed afterwards.

For the assumed circular Fermi pockets, the phase spaces for the BCS (inter-pockets) pairing and the FFLO (intra-pocket) pairing are identical. We proceed to compare the effective pairing interactions in these channels. For clarity, we denote the creation operators for the electronic states close to $\mathbf{K_{1}}$ and $\mathbf{K_{2}}$ by $c^{\dagger}_{\mathbf{k}\uparrow}$ and $\tilde{c}^{\dagger}_{\mathbf{k}\uparrow}$, respectively. Furthermore, we introduce the relative wave vectors as $\mathbf{q}=\mathbf{k}-\mathbf{K_{1}}$ and $\mathbf{q}=\mathbf{k}-\mathbf{K_{2}}$ respectively for states close to  $\mathbf{K_{1}}$ and $\mathbf{K_{2}}$. For small hole doping, $\mathbf{q}$ is also small and we can expand the pairing potential into polynomials of $q_{x}$ and $q_{y}$.

For the FFLO channel, the pairing interaction is expanded approximately by keeping only the leading-order terms as
\begin{eqnarray}
&&H^{FFLO}_{p}=\frac{V^{0}_{p}}{\Omega}\sum\limits_{\mathbf{q},\mathbf{q}'}(q_{x}q_{x}'+q_{y}q_{y}')\cdot  \notag \\
&&\cdot(c^{\dagger}_{\mathbf{q}\uparrow}c^{\dagger}_{-\mathbf{q}\uparrow}c_{-\mathbf{q}'\uparrow}c_{\mathbf{q}'\uparrow} +\tilde{c}^{\dagger}_{\mathbf{q}\uparrow}\tilde{c}^{\dagger}_{-\mathbf{q}\uparrow}\tilde{c}_{-\mathbf{q}'\uparrow}\tilde{c}_{\mathbf{q}'\uparrow}).
\end{eqnarray}
Except for the replacement of $\mathbf{k}$ by $\mathbf{q}$, the above pairing potential is identical to those in Eqs. (1) and (3) for the electron-doped case. Following the same analysis as that for the electron-doped case, we thus expect that an isotropic $p+ip$ (or, put it in another equivalent form, $q+iq$) superconductivity should originate in the two hole-Fermi pockets separated by $2Q$ ($Q=|\mathbf{K_{1}}|$), as the leading FFLO pairing instability.

For the BCS channel, the pairing interaction is expanded approximately by keeping only the leading-order terms as
\begin{equation}
H^{BCS}_{p}=\frac{2V^{0}_{p}}{\Omega}\sum\limits_{\mathbf{q},\mathbf{q}'}(q_{x}q_{x}'+q_{y}q_{y}') c^{\dagger}_{\mathbf{q}\uparrow}\tilde{c}^{\dagger}_{-\mathbf{q}\uparrow}\tilde{c}_{-\mathbf{q}'\uparrow}c_{\mathbf{q}'\uparrow}.
\end{equation}
Interestingly, the approximate pairing interaction for the BCS channel also has the same dependence on the relative wave vectors $\mathbf{q}$ and $\mathbf{q}'$ as the dependence on $\mathbf{k}$ and $\mathbf{k}'$ for the electron-doped case (Eq.(3)). Thus, following the same analysis as that for the electron-doped case, we expect that an isotropic $q+iq$ superconductivity is the leading pairing instability in the BCS pairing channel.

Note that, the approximate pairing interactions in Eqs.(12) and (13) do not depend on which pair of $K$ points we focus on to define the two Fermi pockets. To determine the leading pairing instability for the hole-doped C$_{6}$H$_{1}$ (and also for C$_{6}$H$_{5}$), we have performed mean-field analysis parallel to those done for the electron-doped case. The pairing operator for the leading BCS pairing is taken as $P_{BCS}^{\dagger}(\mathbf{q})=(q_{x}+iq_{y})c^{\dagger}_{\mathbf{q}\uparrow}\tilde{c}^{\dagger}_{-\mathbf{q}\uparrow}$. For the leading FFLO pairing, we define the pairing operator as $P_{FFLO1}^{\dagger}(\mathbf{q})=(q_{x}+iq_{y})c^{\dagger}_{\mathbf{q}\uparrow}c^{\dagger}_{-\mathbf{q}\uparrow}$ for the Fermi pocket surrounding $\mathbf{K}_{1}$ and $P_{FFLO2}^{\dagger}(\mathbf{q})=(q_{x}+iq_{y})\tilde{c}^{\dagger}_{\mathbf{q}\uparrow}\tilde{c}^{\dagger}_{-\mathbf{q}\uparrow}$ for the Fermi pocket surrounding $\mathbf{K}_{2}$.

Under the assumption that the Fermi pockets are ideally circular, the leading BCS and FFLO pairings defined above turn out to share the same transition temperature and ground state energy and are thus completely degenerate. However, as has been mentioned earlier, the Fermi pockets around every $K$ point are in fact noncircular. Similar to the Fermi pockets of graphene, they in fact show a three-fold symmetry with respect to the $K$ point. On the other hand, the two Fermi pockets centering around $\mathbf{K_{1}}$ and $\mathbf{K_{2}}$ are symmetric with respect to the BZ center. As a result of this actual symmetry of the Fermi surface, the phase space for the FFLO pairing is suppressed as compared to that for the BCS pairing. Therefore, for the hole-doped C$_{6}$H$_{1}$ and also for C$_{6}$H$_{5}$, the leading pairing instability is in the BCS channel with $q+iq$ (or equivalently, $p+ip$) symmetry.

To sum up, we have analyzed the leading pairing instability in both electron-doped and hole-doped C$_{6}$H$_{1}$ and C$_{6}$H$_{5}$, in terms of a phenomenological pairing interaction respecting the $C_{3v}$ symmetry of the material and completely spin polarized Fermi surfaces. By comparing the mean-field ground state energies, the chiral $p+ip$ pairing is found to be the leading pairing instability for electron-doped systems. In the hole-doped cases, both FFLO (intra-pocket) pairing and BCS (inter-pockets) pairing are possible. While the two are degenerate for circular Fermi pockets, the BCS pairing is favored by the actual Fermi pockets which has only three-fold symmetry. The leading BCS pairing for the hole-doped C$_{6}$H$_{1}$ has a $p+ip$ symmetry and is thus also chiral. Since the predicted stable pairings in both electron-doped and hole-doped cases support Majorana fermions as edge states and vortex zero modes, the predicted superconducting phases for C$_{6}$H$_{1}$ and C$_{6}$H$_{5}$ may provide new platforms for studying the novel physics in spintronics \cite{spin} and topological quantum computations \cite{tqc}.

%\newpage %Just because of unusual number of tables stacked at end

\end{document}